\documentclass[aps,prl,twocolumn,showpacs,preprintnumbers,amsmath,amssymb]{revtex4}

\usepackage{graphicx}
\usepackage{dcolumn}
\usepackage{bm}

\begin{document}

\preprint{}

\title{Landau level spectroscopy of ultrathin graphite layers}

\author{M.L. Sadowski}
\author{G. Martinez}
\author{M. Potemski}
\affiliation{Grenoble High Magnetic Field Laboratory, CNRS,
Grenoble, France}
\author{C. Berger}
\altaffiliation{also at LEPES, CNRS Grenoble, France}
\author{W.A. de Heer}
\affiliation{Georgia Institute of Technology, Atlanta, Georgia,
USA}

\date{\today}

\begin{abstract}
Far infrared transmission experiments are performed on ultrathin
epitaxial graphite samples in a magnetic field. The observed
cyclotron resonance-like and electron-positron-like transitions
are in excellent agreement with the expectations of a
single-particle model of Dirac fermions in graphene, with an
effective velocity of $\tilde{c}$ = 1.03$\times$10$^6$m/s.
\end{abstract}


\maketitle

The electronic properties of graphite have recently become the
center of considerable attention, following experiments on
graphite monolayers (graphene) \cite{NovoselovSci} and epitaxial
graphene \cite{Berger2}, which led to the discovery of an unusual
sequence of quantum Hall effect states \cite{NovoselovNat,
ZhangNat} and an energy-dependent mass. The considerable interest
in two-dimensional graphite is fuelled by its particular band
structure and ensuing dispersion relation for electrons, leading
to numerous differences with respect to ``conventional"
two-dimensional electron systems (2DES) \cite{Wallace, McClure56,
Haldane, AndoPR02, Gusynin-PRL, Gusynin-microwave, Peres, Berger1,
Berger2}. The band structure of graphene is considered to be
composed of cones located at two inequivalent Brillouin zone
corners at which the conduction and valence bands merge. In the
vicinity of these points the electron energy depends linearly on
its momentum: $E(\overrightarrow{p}) =
\pm\tilde{c}|\overrightarrow{p}|$, which implies that free charge
carriers in graphene are governed not by Schr\"{o}dinger's
equation, but rather by Dirac's equation for zero rest mass
particles, with an effective velocity $\tilde{c}$, which replaces
the speed of light. With the application of an external magnetic
field, the Dirac energy spectrum evolves into Landau levels with
energies given by
\begin{equation}
\label{eq_levels} E_n = sgn(n) \tilde{c}\sqrt{2e\hbar B |n|} =
sgn(n)E_1 \sqrt{|n|}
\end{equation}
where n scans all positive (for electrons) and negative (for
holes) integers and - very importantly - zero. $E_1$ may be
understood as a characteristic energy introduced by the magnetic
field. The square root dependence on B and Landau level index n is
in stark contrast to ``conventional" 2D electrons, where $E_n =
(n+\frac{1}{2})\hbar eB/m, (n\geq 0)$, and the Landau levels are
equally spaced.

The unusual sequence of quantum Hall effect states and an
energy-dependent electron effective mass
\cite{NovoselovNat,ZhangNat}, found in magneto-resistance
measurements, are consistent with the model of Dirac particles.
Here we report a magneto-spectroscopy study of the optical
properties of ultrathin epitaxial graphite layers, in which we
directly probe the dependence of the energy of electrons on their
momentum.

The experiments were performed on graphene layers grown in vacuum
by the thermal decomposition method \cite{Berger1, Berger2}, on
single crystal (4H) SiC. These epitaxial graphene structures are
routinely characterized using low energy electron diffraction,
Auger electron spectroscopy, X-ray diffraction, scanning
tunnelling microscopy and atomic force microscopy. The results of
these measurements in combination with angular resolved
photoelectron spectroscopy and transport data indicate that the
graphitized part of this type of structure consists of a few (3-5)
graphene layers \cite{Berger1, Berger2}. We investigated two such
(unpatterned) structures, with dimensions of about 4 x 4 mm$^2$,
which both show a similar behavior.

The far infra-red transmission of the samples was measured, at a
temperature of 1.9 K, as a function of the magnetic field B. A Si
bolometer was placed directly beneath the sample to detect the
transmitted radiation. The light (provided and analyzed by a
Fourier transform spectrometer) was delivered to the sample by
means of light-pipe optics. All experiments were performed with
non-polarized light, in the Faraday geometry with the wave vector
of the incoming light parallel to the magnetic field direction.
The transmission spectra were normalized by the transmission of
the substrate and by the zero-field transmission, thus correcting
for magnetic field induced variations in the response of the
bolometer. The SiC substrate used was completely opaque for
energies between 85 meV and about 200 meV, which limited the range
of our investigation.

\begin{figure}
\includegraphics{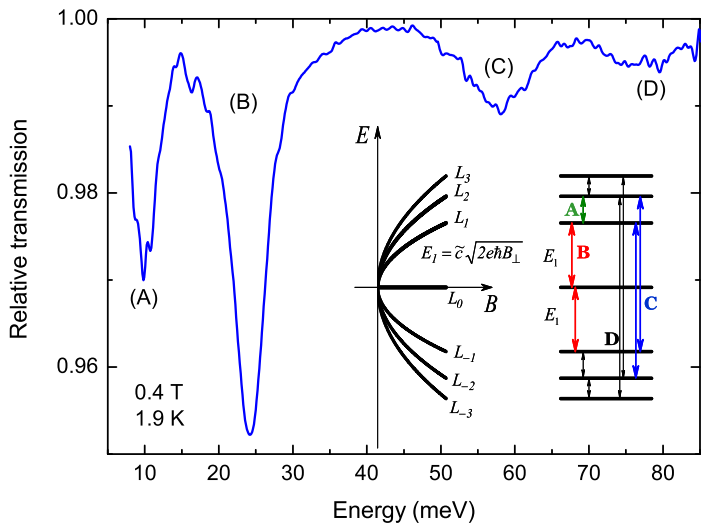}
\caption{\label{fig-transitions} (Color online) Relative
transmission trace at 0.4 T and 1.9 K shows 4 distinct
transitions. The assignations are (see text) A: $L_1 \rightarrow
L_2$, B: $L_0 \rightarrow L_1 (L_{-1} \rightarrow L_0)$, C :
$L_{-2} \rightarrow L_{1} (L_{-1} \rightarrow L_2)$, D: $L_{-3}
\rightarrow L_2 (L_{-2} \rightarrow L_3)$. The inset shows a
schematic of the evolution of Landau levels with applied magnetic
field, and possible optical transitions.}
\end{figure}

\begin{figure}
\includegraphics{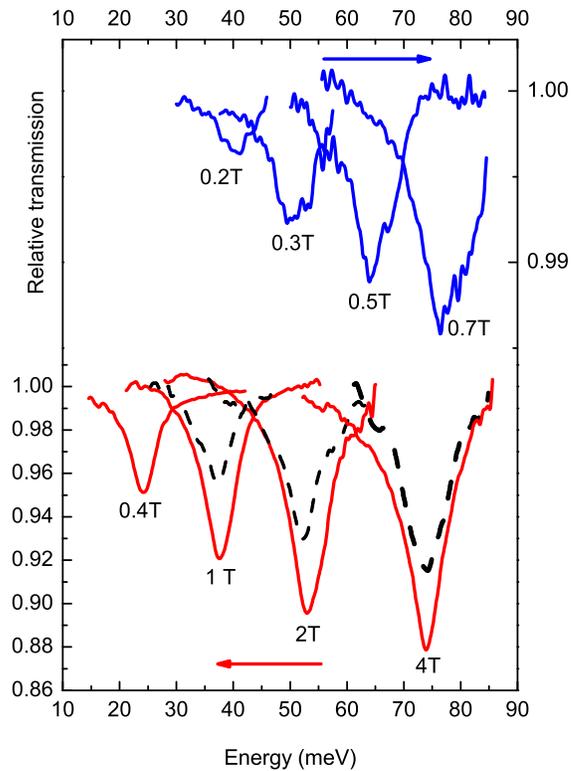}
\caption{\label{fig-spectra} (Color online) Development of two
main transmission lines with the magnetic field. Upper panel shows
the line marked C in Fig.\ref{fig-transitions}, corresponding to
$L_{-1} \rightarrow L_2 (L_{-2} \rightarrow L_1)$ transitions,
lower panel shows the line marked B, corresponding to $L_{0}
\rightarrow L_1 (L_{-1} \rightarrow L_0)$ transitions: solid lines
are for sample 1, dashed lines for sample 2.}
\end{figure}

\begin{figure}
\includegraphics{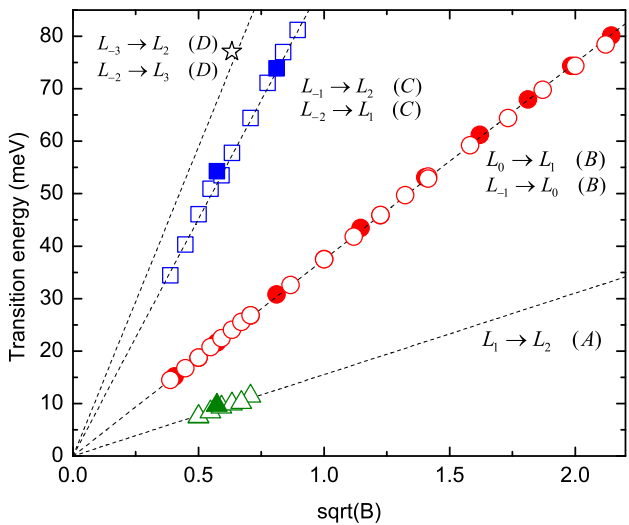}
\caption {\label{fig-linefit} The observed transitions, together
with their assignments, plotted versus $B_\bot$. The filled
symbols are data obtained for the sample tilted with respect to
the direction of B by an angle of 50$^0$. The dashed lines are all
calculated with the same characteristic velocity $\tilde{c} = 1.03
\times 10^{6}$ m/s.}
\end{figure}

The main experimental finding consists of several absorption lines
visible in the spectra. A representative transmission spectrum for
sample 1, at 0.4 T, is shown in Fig.~\ref{fig-transitions}. These
lines evolve spectacularly with the magnetic field. Two main lines
are shown in Fig.~\ref{fig-spectra} for several values of the
field. As shown in Fig.~\ref{fig-linefit}, their energies, plotted
as a function of the square root of the magnetic field, trace
perfect straight lines, in excellent agreement with
eq.~\ref{eq_levels}. Also shown in this figure are the energy
positions of the two other lines. Experiments performed in a
tilted configuration show that the position of the transition line
(filled symbols in Fig.~\ref{fig-linefit}) depends only on the
component of the magnetic field perpendicular to the sample plane.
A complicated structure at still lower energies, not shown in the
figure, moving slowly to higher energies with magnetic field, was
also observed. Unusually, the intensities of the two main lines
increase markedly with increasing magnetic field, with the
lower-energy line always remaining stronger.

Cyclotron resonance in graphite has been studied experimentally
\cite{Galt56} and theoretically \cite{Nozieres58, Inoue62}. These
experiments showed a linear dependence of the cyclotron frequency
on the magnetic field, with an effective mass of 0.058m$_0$. Our
results are best described using the predictions of a simple
single-particle (Dirac) model for a graphene layer, and we will
use this language in the following paragraphs. To facilitate
discussion, we sketch the graphene Landau levels and possible
transitions between them in the inset to
Fig.~\ref{fig-transitions}.

Thus, we assign the strongest line to the transitions to and from
the lowest Landau level. Note that, since the conduction and
valence band states in graphene are built from the same atomic
orbitals, the positive and negative branches of the dispersion
relation are identical. The ensuing symmetry means that the $L_0
\rightarrow L_1$ and $L_{-1} \rightarrow L_0$ transitions are
indistinguishable in an experiment using unpolarized radiation. A
straight line fit of the points corresponding to this transition
using the expression $E = E_1 = \tilde{c} \sqrt{2e\hbar B}$ yields
a very accurate value for $\tilde{c}$, the velocity of electrons
in graphene. This is found for both samples to be (1.03 $\pm$0.01)
$\times 10^6$m/s, consistently with transport measurements
\cite{NovoselovNat,ZhangNat}.

The slopes of the other lines traced in Fig.~\ref{fig-linefit},
starting from the highest energy transition, scale exactly as
$(\sqrt{3}+\sqrt{2}):(\sqrt{2}+1):1:(\sqrt{2}-1)$, allowing these
lines to be assigned to transitions $L_{-2} \rightarrow L_3
(L_{-3} \rightarrow L_2)$, $L_{-1} \rightarrow L_2 (L_{-2}
\rightarrow L_1)$, $L_0 \rightarrow L_1 (L_{-1} \rightarrow L_0)$,
and $L_1 \rightarrow L_2$, respectively, as shown in the figure.

The fact that transitions involving the $L_0$ Landau level are
visible at such low magnetic fields places an upper limit on the
electron concentration in the observed layer. The observation of
the $L_0 \rightarrow L_1 (L_{-1} \rightarrow L_0)$ line implies
the existence of unpopulated states at least on the $L_1$ level.
This line is clearly observed at fields $B \approx 0.15$ T, and
therefore the $L_1$ level can be fully populated only when $B <
0.15$ T. Thus $n \leq 2.1 \times 10^{10}$ cm$^{-2}$ (where we take
into account the 2- and 4-fold degeneracy of the $L_0$ and $L_1$
electronic Landau levels, respectively). This is also consistent
with the disappearance of line A ($L_1 \rightarrow L_2$) when the
$L_1$ level is depopulated by the magnetic field (see
Fig.~\ref{fig-linefit}).

\begin{figure}
\includegraphics{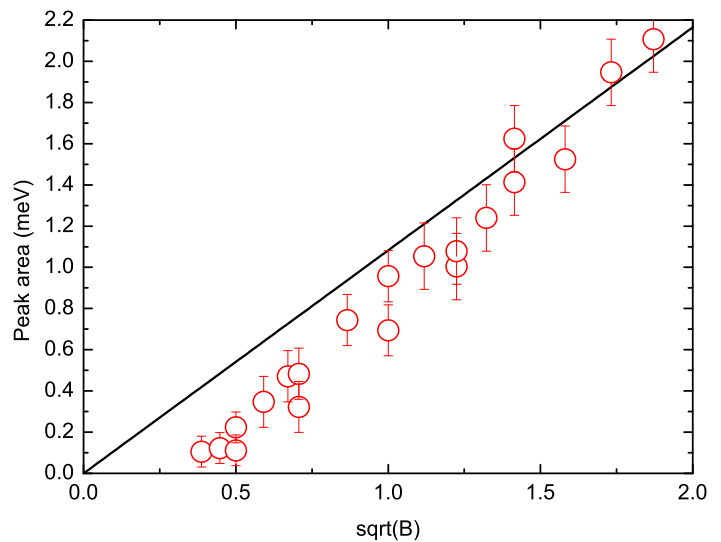}
\caption {\label{fig-oscstrengths} Area of the main peak observed
in the experiments - the $L_0 \rightarrow L_1 (L_{-1} \rightarrow
L_0)$ transition - plotted against the square root of the magnetic
field. The line is traced using the expression
$\beta(e^2/2\epsilon_0\hbar c)E_1$.}
\end{figure}

We now turn our attention to the strength of the transitions. As
may be seen in Fig.~\ref{fig-spectra}, both the main transitions
gain in intensity with increasing magnetic field. To better
visualize this  trend, we plot the integrated intensity(area under
the dip in the relative transmission) for sample 1, as a function
of the square root of the magnetic field
(Fig.~\ref{fig-oscstrengths}). Sample 2 shows the same behavior
but smaller values of the intensity.

The relative transmission of a sheet of conducting electrons
between vacuum and a dispersionless polar medium with a refractive
index $\kappa$, for unpolarized radiation and in the limit of weak
absorption, may be written as (see e.g. \cite{Chiu, AndoCR})
\begin{equation}
\nonumber T(\omega,B) \approx 1 -
\beta\frac{Re(\sigma_{xx}(\omega,B))}{\epsilon_0 c}
\end {equation}
where $\sigma_{xx}(\omega,B)$ is a diagonal element of the optical
conductivity tensor, $\epsilon_0$ is the vacuum permittivity, $c$
is the speed of light in vacuum and $\beta =
(\kappa^2+3)/2(\kappa^2+1)$ = 0.63 for SiC, where $\kappa$=2.6
(from the substrate transmission). The optical conductivity of the
2D electrons may be written using the Kubo formalism \cite{Kubo}
and taking into account the properties of the graphene Landau
level wave functions \cite{AndoPR02, AndoJPSJ02}
\begin{equation}
\nonumber \sigma_{xx}(\omega,B) =  \frac{4G_B
e^2}{\omega}\sum_{m,n} \frac{(f_m -
f_n)M_{m,n}}{E_{m,n}-(\hbar\omega+i\gamma)}
\end{equation}
where $E_{m,n}$ are the transition energies between levels $m$ and
$n$, $G_B = eB/h$ is the Landau level degeneracy, $f_m, f_n$ are
the occupancies of the relevant Landau levels and the selection
rules for the optically active transitions are given by
$M_{m,n}=(\tilde{c}^2/p)\delta_{\mid m\mid,\mid n\mid\pm 1}$, with
p = 2 for m or n =0 and 4 otherwise. The summation is performed
over all Landau levels $m, n$, and the fourfold degeneracy of each
Landau level has already been accounted for.

The integrated transmission for a single transition between a
completely filled ($L_0$) and a completely empty ($L_1$) Landau
level, using the above expression (for linewidths $\gamma \ll
E_{m,n}$) may be written as:
\begin{equation}
\nonumber I(B) =
\frac{1}{\epsilon_0c}\int{Re(\sigma_{xx}(\omega))d\omega} \approx
\frac{e^3\tilde{c}^2B}{\epsilon_0c E_1} =
\beta\frac{e^2\tilde{c}}{2\epsilon_0\hbar c}E_1
\end{equation}
where $E_1$ is the characteristic energy introduced earlier.

The above equation gives a rough estimate of the intensity of the
strongest transition, in the range of high magnetic fields where
the Fermi energy is pinned to the $L_0$ level. This is due to the
fact that the decreasing intensity of the $L_0 \rightarrow L_1$
transition is compensated by the corresponding increase of the
strength of the superimposed $L_{-1} \rightarrow L_0$ transition.
At low magnetic fields, where the $L_1$ level is not completely
empty, the observed oscillator strength decreases, disappearing
when the $L_1$ is fully populated. In spite of the rather crude
approximation, Fig.~\ref{fig-oscstrengths} indeed shows that the
observed transition follows the expected trend; the good agreement
of the absolute measured and calculated values is another factor
supporting the picture of a single, possibly inhomogeneous,
graphene layer (see discussion in following paragraphs).

Several notable differences emerge between Dirac electrons and
conventional two-dimensional electron systems (2DES). As we have
shown, transitions between adjacent Landau levels in graphene
occur at markedly different energies (for example $L_0 \rightarrow
L_1$ and $L_1 \rightarrow L_2$, Figs.~\ref{fig-transitions}
and~\ref{fig-linefit}). For a standard 2DES, transitions between
such pairs of Landau levels all have the same energy. More
striking, a different class of transitions, with no counterpart in
a standard 2DES, is observed in graphene and involves those from
hole ($n < 0$) to electron ($n
> 0$) states (e.g. our $L_{-1} \rightarrow L_2$ and $L_{-2}
\rightarrow L_1$ transitions). These are the particle-antiparticle
creation and annihilation events in the Dirac formalism.

Since some of the observed transitions are analogues of cyclotron
resonance, it is tempting to look at them in a semi-classical
context, using the concept of an effective mass. While for a 2DES
with a quadratic dispersion law there is a coincidence between
classical and quantum mechanical solutions of the optically active
response in a magnetic field, this does not hold for graphene. The
classically derived cyclotron excitation $E_C$ in this system is
$E_C = \hbar eB/(E/\tilde{c}^2)$ \cite{AndoPR02}, where $E$ is the
electron energy and $(E/\tilde{c}^2)$ stands for the electron
mass. Although the effective rest mass of the electrons in
graphene is zero, their energy- and magnetic field-dependent
cyclotron mass can be followed down to the lowest energies
($\simeq 7$ meV in our case), giving a lowest observed value of
0.0012 $m_0$.

Having demonstrated the presence of zero effective rest mass Dirac
fermions in the investigated structure, let us now consider the
following points: (i) linear dispersion is characteristic of a
single graphene layer, while a graphene bilayer \cite{McCann,
NovoselovNatPhys} is found to exhibit parabolic dispersion; (ii)
transport measurements performed on a mesoscopic sample patterned
on the same wafer as our selected sample, which also show the
unusual Berry's phase of $\pi$ observed in graphene, give a
concentration of $\simeq 4 \times 10^{12} cm^{-2}$, while our
results point to a concentration two orders of magnitude smaller.

It is believed that electric transport is dominated by the
interface layer, which has a high electron concentration due to
the built-in electric field caused by the surface charge
\cite{Visscher, Berger2}. Transmission measurements, on the other
hand, probe the whole sequence of layers, including those further
away from the interface which have lower electron concentrations.
It is possible that the observed Dirac spectrum originates from a
single graphene layer ``floating" above a SiC substrate covered
with other graphitic layers \cite{Forbeaux}. The previously
mentioned very low-energy features in our spectra could arise from
the high-electron-concentration parts of the sample, where the
energy difference between adjacent Landau levels is small.

Another factor possibly affecting the data could be lateral
inhomogeneity within a single graphene plane, or even
fragmentation of the layer - this could explain the weaker
intensities observed for sample 2. Individual graphene planes in
epitaxial graphite may be much more weakly coupled than is usually
accepted for graphite. Several graphene layers
\cite{Guinea-multilayer} may, depending on the stacking scheme,
exhibit linear and/or parabolic dispersion relations. Finally, we
note that carriers with linear dispersion may also be found at the
H point of bulk graphite \cite{Toy,Partoens,Zhou}, although the
significant difference between the value of $\tilde{c}$ found in
\cite{Zhou} (0.91 $\times 10^6$ m/s) and our experiment, as well
as the structure of the samples, makes such an interpretation
unlikely. The current experiment shows an absorption in good
agreement with that expected for graphene, but the simple
approximation used does not exclude more complex scenarios.

Concluding, we have measured the optical excitation spectrum of
(relativistic-like) Dirac fermions in a condensed matter system.
These fermions are found in thin layers of epitaxial graphite,
probably in single (or extremely weakly coupled) graphene layers
(or parts of layers). Cyclotron resonance like transitions coexist
with electron-hole (particle-antiparticle) like transitions, with
energy positions and oscillator strengths in surprisingly good
agreement with expectations based on a model of non-interacting
particles with linear dispersion.

The GHMFL is a ``Laboratoire conventionn\'{e} avec l'UJF et l'INPG
de Grenoble". The present work was supported in part by the
European Commission through grant RITA-CT-2003-505474 and by
grants from the Intel Research Corporation and the NSF: NIRT
``Electronic Devices from Nano-Patterned Epitaxial Graphite".

\bibliographystyle{aipproc}

\end{document}